\documentclass[12pt]{amsart}
\usepackage{amsmath}
\newcommand{\tc}{\widetilde{\mathcal C}}
\newcommand{\tp}{\widetilde{P}}
\newcommand{\cal}{\mathcal}

\newcommand{\oa}{(\Omega,\cal A)}
\newcommand{\te}{\widetilde{E}}

\newcommand{\eh}{{\cal E}\,({\cal{H})}}

\newcommand{\lh}{{\cal L}({\cal H})}
\newcommand{\ph}{{\cal P}({\cal H})}

\newcommand{\hi}{{\cal H}}
\newcommand{\lk}{\cal L(\cal K)}
\newcommand{\ki}{\cal K}
\newcommand{\pk}{{\cal P}({\cal K})}
\newcommand{\ip}[2]{\left\langle\,#1\,|\,#2\,\right\rangle}

\newcommand{\fii}{\varphi}

\newtheorem{theorem}{Theorem}%[section]
\newtheorem{lemma}[theorem]{Lemma}%[section]
\newtheorem{remark}[theorem]{Remark}%[section]
\newtheorem{corollary}[theorem]{Corollary}%[section]
\newtheorem{proposition}[theorem]{Proposition}
\begin{document}
\title[Coexistence]{Notes on coarse grainings and functions of  observables}
\author{A. Dvure\v censkij}
\address{Anatolij Dvure\v censkij,
Mathematical Institute, Slovak Academy of Sciences,
SK-81473 Bratislava, Slovakia}
\email{dvurecen@mat.savba.sk}
\author{P. Lahti}
\address{Pekka Lahti,
Department of Physics, University of Turku, 
FIN-20014 Turku, Finland}
\email{pekka.lahti@utu.fi}
\author{S. Pulmannov\' a}
\address{Sylvia Pulmannov\' a,
Mathematical Institute, Slovak Academy of Sciences,
SK-81473 Bratislava, Slovakia}
\email{pulmann@mat.savba.sk}
\author{K. Ylinen}
\address{Kari Ylinen,
Department of Mathematics, University of Turku, 
FIN-20014 Turku, Finland}
\email{kari.ylinen@utu.fi}

%\date{\today}
\begin{abstract}
Using the Naimark dilation theory we investigate the question under what conditions
an observable %(semisepctral measure)  $E_1$ 
which is a coarse graining of another
observable %(semispectral measure) $E$ 
is a function of it.  To this end, conditions for the separability and for the Boolean structure of an 
observable are given.


\end{abstract}

\maketitle

\section{Intoduction}

Let $\oa$ be a measurable space, $\hi$ a complex %separable 
Hilbert space, $\lh$ the set of bounded operators on $\hi$, and
$E:\cal A\to\lh$
a normalized positive operator measure, that is, 
a semispectral measure. %POM for short. 
We call such measures {\em observables} of a physical system described by $\hi$.

Let $(\ki,\te,V)$ be a Naimark dilation of $E$ into a spectral measure $\te$,
that is, $\te:\cal A\to\lk$ is a projection measure acting on a Hilbert space $\ki$
and 
$V:\hi\to\ki$  an isometric linear map such that 
$E(X) = V^*\te(X)V$ for all $X\in\cal A$.
We say that an observable $E:\cal A\to\lh$ is {\em separable} 
if it has a Naimark dilation $(\ki,\te,V)$ 
which is separable, that is, the range  $\te(\cal A)$ of $\te$ 
is a separable Boolean sub-$\sigma$-algebra in the projection lattice $\pk$
of the Hilbert space $\ki$.  (We use the lattice theoretical terminology as introduced in \cite{VSV85}.)

We recall that a Boolean sub-$\sigma$-algebra $\cal B$ of $\pk$ 
is  separable, 
if there exists a countable subset $B$  such that the smallest Boolean sub-$\sigma$-algebra of  $\cal B$
containg $B$  is $\cal B$. % \cite[p. 15]{VSV85}. 
The importance of such sub-$\sigma$-algebras of $\pk$ lies in the following fact:
a Boolean  sub-$\sigma$-algebra $\cal R$ of $\pk$ is the range of a real projection measure $F:\cal B(\mathbb R)\to\lk$,
that is, $\cal R= F(\cal B(\mathbb R))$ if and only if $\cal R$ is separable \cite[Lemma 3.16]{VSV85}.
Furthermore, in that case, if $\cal R_1$ is a Boolean sub-$\sigma$-algebra contained in $\cal R$, 
%which is separable, 
then there is a real Borel function
$f$ such that $\cal R_1=F^f(\cal B(\mathbb R))$, where $F^f(X)=F(f^{-1}(X))$ \cite[Theorem 3.9]{VSV85}, see also
\cite[Lemma 4.11]{Gudder70}.

Consider now two observables $E_1$ and $E$ defined on the $\sigma$-algebras $\cal A_1$ and $\cal A$ of the
measurable spaces $(\Omega_1,\cal A_1)$
and $\oa$, respectively, and taking values in $\lh$.
We say that $E_1$ is a {\em function} of $E$ if there is a measurable function $f:\Omega\to\Omega_1$ such that
$E_1= E^f$, that is, $E_1(X)=E(f^{-1}(X))$ for all $X\in\mathcal A_1$.
Clearly, if $E_1$ is a function of $E$, then the range
%$E_1(\cal A_1)$ 
of $E_1$ is
contained in the range  
%$E(\cal A)$ 
of $E$.
In general,  for any two observables $E_1$ and $E$,  if
$E_1(\cal A_1)\subset E(\cal A)$ we say that $E_1$ is a {\em coarse graining} of $E$.

Assume that $E_1$ is a coarse graining of $E$.
If  $(\ki,\te,V)$
is a Naimark dilation of $E$, 
we let  $\cal R_1$ be the set of  all projections $P\in\te(\cal A)$ 
such that 
$V^*PV\in E_1(\cal A_1)$.
Then
$$
E_1(\cal A_1)= V^*\cal R_1V \subset 
E(\cal A)=V^*\te(\cal A)V.
$$ 
Calling two observables  {\em equivalent} if their ranges are the same we observe that if $\te({\cal A})$ is a separable 
Boolean sub-$\sigma$-algebra of $\pk$, then $\te$ is equivalent to a real projection measure
$F:\cal B(\mathbb R)\to\lk$. If, in addition, $\cal R_1$ is  a Boolean sub-$\sigma$-algebra of $\te(\cal A)$
then it can be expressed as $\cal R_1=F^f(\cal B(\mathbb R))$ for some Borel  function $f$. 
In this case observables $E_1$ and $E$ are equivalent to the two real functionally related semispectral measures $E_1^r$ and $E^r$, where
$E_1^r(X)= V^*F^f(X)V$ and $E^r(X)=V^*F(X)V$ for all $X\in\cal B(\mathbb R)$.

The questions of interest for this study are  the following. 
First, under what conditions is an observable  separable?
Secondly, if an observable % $E_1$ 
is a coarse graining of 
another observable, %$E$, 
when is it %$E_1$ 
a function of the latter? % $E$? 
Sections~\ref{sigma} and \ref{function} are devoted to the separability questions 
whereas in Section~\ref{Boolean} we study  
the question of functional relations between observables.

\begin{remark}\label{huomautus}{\rm 
For positive operator measures $E_1$ and $E$, the condition $E_1(\cal A_1)\subset E(\cal A)$ need not imply that $E_1$ is a
function of $E$.  
However, $E_1$ and $E$ may still be functionally related 
(functionally coexistent)
so that there is a positive operator measure
$F$ with measurable functions $f$ and $g$ such that $E_1=F\circ f^{-1}$ and $E=F\circ g^{-1}$.
Indeed, as an illustration of this phenomenon, consider the real scalar measures
$E$ and $E_1$ 
 concentrated, respectively, on the sets $\{x_1,x_2,x_3,x_4\}$
and $\{y_1,y_2,y_3,y_4\}$ such that 
$ E(\{x_1\})=E(\{x_2\})=1/8, E(\{x_3\})=E(\{x_4\})=3/8$, 
and
$ E_1(\{y_1\})=E_1(\{y_2\})=E_1(\{y_3\})=1/8, E_1(\{y_4\})=5/8$. 
Clearly, the range of 
$E_1$ is contained in that of $E$, but there is no function 
$f:\{x_1,x_2,x_3,x_4\}\to \{y_1,y_2,y_3,y_4\}$  such that
$E_1(Y)=E(f^{-1}(Y))$. Indeed, if such a function exists, we must have 
$E_1(\{y_1\})=E( f^{-1}(\{y_1\}))=1/8$, 
which gives $f^{-1}(\{y_1\})=\{x_1\}$, or 
$f^{-1}(\{y_1\})=\{x_2\}$, 
and
$E_1(\{y_4\})=E(f^{-1}(\{y_4\}))$, 
which yields 
$f^{-1}(\{y_4\})=\{x_1,x_2,x_3\}$
 or $f^{-1}(\{y_4\})=\{x_1,x_2,x_4\}$.
Both $E$ and $E_1$ are, however,  functions of 
the %eight-valued 
observable $\{z_i\}\mapsto F(\{z_i\})=1/8$, $i=1,\ldots,8$.
}\end{remark}

\section{Separable Boolean $\sigma$-algebras}\label{sigma}
In this section we collect, for the  reader's convenience, some basic observations in the context
of  separable Boolean sub-$\sigma$-algebras of the projection lattice of a
Hilbert space 
The proofs follow readily from known facts
and the results themselves may be part of the folklore of the subject though hard to find
in the literature.

Let $\cal B$ be a Boolean algebra. An {\it atom} of $\cal B$ is any non-zero element $a$ of 
 $\cal B$ such that $b \le a$ for $b \in \cal B$ implies $b = 0$ or $b = a.$ Let $\mbox{At}(\cal B)$ be the
 set of all atoms of $\cal B$. If $\mbox{At}(\cal B)=\emptyset$, $\cal B$ is said to be {\it atomless}. 
If $a$ and $b$ are two different atoms of $\cal B$, then they are disjoint, $a \wedge b = 0.$

 If $ \cal B_i=(\cal B_i;0_i,1_i,^{'i})$, $i=1,2,$ are Boolean $\sigma$-algebras, then their 
 Cartesian product $\cal B = \cal B_1 \times \cal B_2$ is again a Boolean $\sigma$-algebra
 with operations defined  coordinatewise, the least and the greatest elements being $0=(0_1,0_2)$ and $1=(1_1,1_2)$, respectively.

 \begin{proposition}\label{pr:1.1} 
Let $\cal B$ be a Boolean $\sigma$-algebra such 
 that every system of mutually orthogonal non-zero elements of $\cal B$ is at most countable. 
 Then $\cal B$ can be decomposed in the form $\cal B = \cal B_1 \times \cal B_2$, where $\cal B_1$ is a 
 Boolean $\sigma$-algebra isomorphic with the power set $2^N,$ where $N$ is a finite or 
 countable cardinal, and $\cal B_2$ is an atomless Boolean $\sigma$-algebra.
 \end{proposition}

 \begin{proof} 
Let $\mbox{At}(\cal B)$ be the set of all atoms of $\cal B$. Since any two 
 different  atoms $a$ and $b$ of $\cal B$ are mutually orthogonal, $a\leq b'$, 
$0 \le |\mbox{At}(\cal B)| \le\aleph_0.$

 Define $a_0 := \bigvee\{a:\ a \in \mbox{At}(\cal B)\}$; if $\mbox{At}(\cal B)= \emptyset$,
 we put $a_0 :=0$. For any element $a \in\cal B $, we have the decomposition
 $$
 a = (a \wedge a_0) \vee (a\wedge a_0').\eqno(1.1)
 $$                              
 Define $\cal B_1 :=\{a \in \cal B:\ a \le a_0\}$ and $\cal B_2 :=  \{a \in \cal B:\ a \le a_0'\}$.
 Then $\cal B_1 =(\cal B_1;0,a_0,^{'a_0})$, where $x^{'a_0} := x' \wedge a_0$ for
 $x \in \cal B_1$, and $\cal B_2 =(\cal B_2;0,a_0',^{'a_0'})$, where $x^{'a_0'} :=x' \wedge a_0' $ for
 $x \in \cal B_2$,  are Boolean $\sigma$-algebras such that $\cal B_1$ is isomorphic
 with the $\sigma$-algebra $2^N$, where $N= |\mbox{At}(\cal B)|$, and $\cal B_2$ is atomless.
 In view of (1.1)  we have the decomposition $\cal B= \cal B_1\times \cal B_2.$
 %\hfill{$\Box$}
\end{proof}

 \vspace{3mm}
 The set ${\cal P}(\hi)$  of all projections on $\hi$  
 forms  a complete orthomodular lattice with respect to 
the operator order  and  orthocomplementation 
$P\mapsto P^\perp := I_\hi-P$, with $I_\hi=I$ 
%is the identity  operators  $\hi$. 
and $O_\hi=O$ being the identity and zero operators on $\hi$. 

 \begin{theorem}\label{th:1.2} 
Let $\hi$ be  a complex separable Hilbert space and
let $\cal B$ be a Boolean sub-$\sigma$-algebra of  ${\cal P}(\hi)$.  Then $\cal B$ is separable. 
 In particular, if $\hi$ is  finite dimensional, 
 then $\cal B = 2^N$, where $N$ is an integer such that $1\le N \le \dim\hi.$
 \end{theorem}

 \begin{proof} Using Proposition \ref  {pr:1.1} we decompose the 
 $\sigma$-algebra $\cal B$ in the form $\cal B= \cal B_1 \times \cal B_2$, where $\cal B_1$ is isomorphic
 with $2^N$, $N = |\mbox{At}(\cal B)|$, and $\cal B_2$ is atomless. Let $P_0 = \bigvee
 \{P:\ P\in \mbox{At}(\cal B)\}$ and denote  $\hi_0=P_0(\hi)$.

 Assume $\dim\hi = \aleph_0$. If $P_0 = I_\hi$, then $\cal B=\cal B_1$, and $\cal B$ is separable.
 If $P_0 \ne I_\hi,$ then $I_\hi-P_0 \ne O$, and since $\cal B_2$ is atomless, we have
 $\dim (\hi_0) = \aleph_0.$ In addition, $\cal B_2$ is therefore a Boolean 
 $\sigma$-algebra which is a subalgebra  of ${\cal P}(\hi_0^\bot).$ 
Let ${\mathbb B}_2$
 be the von Neumann algebra generated by $\cal B_2$. Then ${\mathbb B}_2$ is a commutative
 von Neumann algebra acting in the infinite-dimensional complex separable 
 Hilbert space $\hi_0^\bot$ and the projection lattice of ${\mathbb B}_2$ coincides
 with $\cal B_2$ which is atomless.
 Therefore, by \cite[Theorem III.1.22]{Takesaki}, ${\mathbb B}_2$ is 
 isomorphic with the von Neumann algebra $L^\infty(0,1) $ (the space of all 
 essentially bounded functions on the unit interval $(0,1)$ with respect to the
 Lebesgue measure). Since the projections from $L^\infty(0,1) $ are only 
 characteristic functions, they have a countable generator, consequently,
 $\cal B_2$  has a countable generator. Because  $\cal B_1$ is generated by the countable
 set of atoms, in view of $\cal B = \cal B_1 \times \cal B_2$, $\cal B$ is separable.

 Assume now $\dim\hi < \infty.$ Then $P_0 =I_\hi$ and therefore, $\cal B = \cal B_1 = 2^N.$
 %\hfill{$\Box$}
\end{proof}

\section{Separable observables}\label{function}

A Naimark dilation $(\ki,\te,V)$ of a semispectral measure $E:\cal A\to\lh$ is {\em minimal}
if $\ki$ is the closed linear span of $\{\te(X)\,|\, X\in\cal A\}$. As is well known, a minimal dilation
always exists and it is unique up to an isometric isomorphism \cite{Paulsen}.

\begin{lemma}\label{lemma_sep}
Let $\oa$ be a measurable space with a separable $\sigma$-algebra $\cal A$
and let $E:\cal A\to\lh$ be a normalized positive operator measure acting on a complex
separable Hilbert space $\hi$. 
If $(\ki,\te,V)$ is a minimal Naimark dilation of $E$, then $\ki$ is separable.
\end{lemma}

\begin{proof}
Let $\mathcal F$ be a countable collection of subsets of $\Omega$ which generates 
the $\sigma$-algebra $\mathcal A$,
%, that is, $\mathcal A= \sigma(\mathcal F)$.
and let $\mathcal R$ be the ring generated by $\mathcal F$. Since $\mathcal F$ is countable,
the ring $\mathcal R$ is countable \cite[Theorem I.5.C]{Halmos}.
Let $\mathcal C$ be the complex linear span of the characteristic functions
$\chi_X$ of the sets $X\in\mathcal R$, and
let $\tc$ be its closure  in the set of bounded functions  $\Omega\to\mathbb C$ (with respect to the sup-norm).
$\tc$ is a separable commutative $C^*$-algebra.
Let $\Phi:\tc\to\lh$ be the positive  linear map corresponding to the normalized positive operator measure $E:\mathcal A\to\lh$,
$\Phi(f)=\int f\,dE$. Then $\Phi$   is completely positive  \cite[Theorem 3.10]{Paulsen}.
%$\Psi:\tc\to\lk$ 
Let $(\ki,\pi,V)$ be its minimal Stinespring dilation. The Hilbert space $\ki$ is separable
\cite[p. 46]{Paulsen}.
Let $P_o:\mathcal R\to\lk$ be the projection-valued set function defined by $P_o(X)=\pi(\chi_X)$ for all $X\in\mathcal R$.
Then $V^*P_o(X)V=E(X)$ for all $X\in\mathcal R$.
From its construction it easily follows that $P_o$ is weakly $\sigma$-additive on $\mathcal R$.

For any $\fii\in\ki$ and $X\in\mathcal R$ denote $\mu^o_{\fii,\fii}(X)=\ip{\fii}{P_o(X)\fii}$.
Since $\mu^o_{\fii,\fii}$ is $\sigma$-additive on $\mathcal R$, it has a unique extension to a (positive) measure
$\mu_{\fii,\fii}$ on $\mathcal A$.
For any $\fii,\psi\in\ki$, we let $\mu_{\fii,\psi}= \frac 14\sum_{k=1}^4i^k \mu_{\fii+i^k\psi,\fii+i^k\psi}$.
Elementary estimates show that the map $(\fii,\psi)\mapsto \mu_{\fii,\psi}(X)$ is a bounded sesquilinear form for each $X\in\mathcal A$,
and we get a positive operator measure $\tp:\mathcal A\to\lk$ which extends $P_o$.
% satifying $\ip{\fii}{\tp(X)\psi}= \mu_{\fii,\psi}(X)$.

It remains to be shown that the  map $\tp$ is a projection measure.
We denote by $M(\mathcal R)$  the monotone class generated by $\mathcal R$.
The class  $\{X\in\mathcal A\,|\, \tp(X)^2=\tp(X)\}$
contains $\mathcal R$ and is easily seen to be a monotone class and so it equals $\mathcal A$  \cite[Theorem I.6.B]{Halmos}.
Clearly,  $V^*\tp(X)V=E(X)$ for all $X\in\mathcal A$ and 
$(\ki,\tp,V)$ constitutes a minimal dilation of $E$ and $\ki$ is separable.
\end{proof}

An alternative approach would be to use in the above proof  
Naimark's dilation theory  \cite[Appendix, Theorem 1]{Nagy}
instead of Stinespring's.

\begin{remark}{\rm
A physically relevant dilation $(\ki,\te,V)$ of a quantum observable $E$ is typically not minimal, see e.g. \cite{LY}.
An interesting example of a dilation acting on a nonseparable Hilbert space appears in \cite{Ozawa} for the 
canonical phase observable.
}\end{remark}

\begin{corollary}
Let $\oa$ be a measurable space with a separable $\sigma$-algebra $\cal A$
and let
$\hi$ be  a complex  separable Hilbert space. Any
normalized positive operator measure $E:\cal A\to\lh$ is separable.
\end{corollary}

\begin{proof}
Let $(\ki,\te,V)$ constitute a minimal Naimark dilation of $E$. The set $\{\te(X)\fii\,|\, X\in\cal A,\fii\in\hi\}$
is dense in $\ki$. 
By Lemma~\ref{lemma_sep} $\ki$ is separable.
Therefore,
by Theorem~\ref{th:1.2}
$\te(\cal A)$ is a separable Boolean sub-$\sigma$-algebra of $\pk$. 
\end{proof}

\section{Boolean observables}\label{Boolean}

The Boolean structure of the range of an observable plays an important 
role in the functional calculus of observables. We therefore recall the following results.
Here  $\eh$  denotes the set of effect operators on $\hi$, i.e.,
 $\eh =\{A \in\lh:\ O \le A \le I\}$.

\begin{proposition}
The range $E(\cal A)$ of an observable $E:\cal A\to\lh$ is a Boolean subalgebra of the set $\eh$ of effects if and only if $E$ is projection valued.
\end{proposition}

\begin{proof} 
For any $X\in\cal A$ the product $E(X)E(X')$ is a positive lower bound of $E(X)$ and $E(X')$. 
If $E(\cal A)$ is Boolean then $E(X)\land E(X')=O$, and thus
$E(X)E(X')=O$, that is, $E(X)^2=E(X)$. 
On the other hand, if $E$ is projection valued, then the claim follows from the multiplicativity of the spectral measure and 
from the fact that for any two projections $P$ and $R$ their greatest lower bound and smallest upper bound in $\eh$ are 
the same as in $\ph$, that is, $P\land R$ and $P\lor R$, respectively.
\end{proof}

The order structure of the set of effects $\eh$ is highly complicated. 
For instance, 
if $E:\cal A\to\lh$ is an observable, then for any $X,Y\in\cal A$, 
the effect $E(X\cap Y)$ is a lower bound of the effects $E(X)$ and $E(Y)$, 
but these effects need not have
the greatest lower bound $E(X)\wedge_{\eh}E(Y)$ and even if $E(X)\wedge_{\eh}E(Y)$ 
exists it need not coincide with $E(X\cap Y)$.
When the order and the complement of $\eh$ are restricted to the range $E(\cal A)$ of $E$ it is
possible that the system $(E(\cal A),\leq,')$ is a Boolean $\sigma$-algebra without $E$ being projection valued.
To express that option it is useful to introduce
two further concepts. We say that an observable $E:\cal A\to\lh$ is {\em regular} 
if for any  $O\ne E(X)\ne I$, neither $E(X)\leq E(X')$ nor $E(X')\leq E(X)$, and  it is {\em $\Delta$-closed} if for
any triple of pairwise orthogonal elements $A,B,C\in E(\cal A)$, the sum %s $A+B$, $A+C$, $B+C$, and 
$A+B+C$
is in $E(\cal A)$. From  \cite{PLMM92,DP94,PLMM95} the following results are then obtained.

\begin{proposition}
For any  observable $E:\cal A\to\lh$ 
the following three conditions are equivalent.
\begin{itemize}
\item[$a)$] $(E(\cal A),\leq,')$ is a Boolean $\sigma$-algebra.
\item[$b)$] $E$ is regular.
\item[$c)$] $E$ is $\Delta$-closed.
\end{itemize}
\end{proposition}

Consider now two observables $E_1$ and $E$ defined on the $\sigma$-algebras $\cal A_1$ and $\cal A$ of the
measurable spaces $(\Omega_1,\cal A_1)$
and $\oa$, respectively, and taking values in $\lh$, with $\hi$ being complex and separable.
Assume that $E_1$ is a coarse graining of $E$, that is,
$E_1(\cal A_1)\subset E(\cal A)$. 
Let  $(\ki,\te,V)$ be a Naimark dilation of $E$, with   separable $\ki$,
%with $V$ as the associated isometry $\hi\to\ki$, 
and let $\cal R_1$ be again the set of projections  $P\in\te(\cal A)$
such that  $V^*PV\in E_1(\cal A_1)$.

\begin{proposition}
With the above notations,
$\cal R_1$ is a Boolean sub-$\sigma$-algebra of $\pk$ if and only if  there is a real Borel function $f$ and a real semispectral measure $E_r$
such that $E$ is equivalent with $E_r$ and $E_1$ is equivalent with $E^f_r$.
\end{proposition}

\begin{proof}
If $\cal R_1$ is a Boolean sub-$\sigma$-algebra of $\pk$  then, as a subset  of $\te(\cal A)$, it  is also separable.
Thus by the results  \cite[Lemma 3.16, Theorem 3.9]{VSV85} %and \cite[Theorem 3.9]{VSV85} 
there is a real
projection measure $F_r$  and a real Borel function $f$ such that 
$\te(\cal A)= F_r(\cal B(\mathbb R))$ and $\cal R_1=F^f_r(\cal B(\mathbb R))$. The semispectral measures
$E_r:=V^*F_rV$ and $E^f_r:=V^*F^f_rV$ are now as required. The other direction is immediate.
\end{proof}

We say that an observable $E:\cal A\to\lh$ has the  {\em V-property} with respect to a subset $Q$ of $E(\cal A)$ if for each $X,Y\in\cal A$
and $C\in Q$ the inequality $E(X)\leq C\leq E(Y)$ implies that there is a $Z\in\cal A$ such that $X\subset Z\subset Y$ and $C=E(Z)$.
The importance of this property  is in the fact that for any two (real) observables $E_1$ and $E$, if $E_1(\cal A)\subset E(\cal A)$ and if
$E$ has the $V$-property on $E_1(\cal A)$, then $E_1$ is a function of $E$ \cite{PLSP}.

\begin{lemma}
With the above notations, $O_\ki, I_\ki\in\cal R_1$, and if $P\in\cal R_1$  then also $P^\perp\in\cal R_1$.
Moreover, for any $P,R\in\cal R_1$, if $P\leq R$, then $V^*PV\leq V^*RV$.
In addition, the  observable $\te$ has the $V$-propety on $\mathcal R_1$.
\end{lemma}
\begin{proof}
If $P\in\cal R_1$, then $V^*PV=E_1(X)$ for some $X\in\cal A_1$ and thus 
$E_1(X')=I_\hi-E_1(X)=V^*V-V^*PV=V^*(I_\ki-P)V$, so that $P^\perp\in\cal R_1$.
If $P\leq R$, then for any $\psi\in\ki$, $\ip{\psi}{P\psi}\leq\ip{\psi}{R\psi}$, and thus, in particular,
for any $\fii\in\hi$, $\ip{\fii}{E_1(X)\fii}=\ip{\fii}{V^*PV\fii}=\ip{V\fii}{PV\fii}\leq\ip{V\fii}{RV\fii}
=\ip{\fii}{V^*RV\fii}=\ip{\fii}{E_1(Y)\fii}$.
To demonstrate the $V$-property, let $X,Y\in\mathcal A$, $X\subseteq Y$, so that $\te(X)\leq\te(Y)$.
Assume that   $P\in\mathcal R_1$ is such that
$\te(X)\leq P\leq\te(Y)$. 
Let $Z\in\mathcal A$ be such that $\te(Z)=P$.  Then for $Z_1=X\cup(Y\cap Z)$ we have $X\subseteq Z_1\subseteq Y$, and 
$\te(Z_1)=\te(X)\lor(\te(Y)\land\te(Z))=(\te(X)\lor\te(Y))\land(\te(X)\lor P)=\te(Y)\land P=P$.
\end{proof}

\begin{remark}{\rm 
The assumption that $\te$ has the $V$-property on $\cal R_1$ does not imply that $E$ has the $V$-property on $E_1(\cal A)$.
For an illustration, see Remark~\ref{huomautus}.
}\end{remark}

\begin{proposition}
With the above notations, if $E_1$ is projection valued, then $\cal R_1$ is a Boolean sub-$\sigma$-algebra of
$\te(\cal A)$.
\end{proposition}

\begin{proof}
For any $P\in\pk$, $V^*PV\in\ph$ if and only if $VV^*P=PVV^*$. 
Let 
$P,R\in\cal R_1$ so that there are
$X,Y\in\cal A_1$ such that $V^*PV=E_1(X)$ and $V^*RV=E_1(Y)$. Then
\begin{eqnarray*}
V^*P\land RV&&=V^*PRV=V^*VV^*PRV\\
&&=V^*PVV^*RV
= E_1(X)E_1(Y)\\
&&=E_1(X\cap Y)
\end{eqnarray*}
showing that $\cal R_1$ is closed under $\land$. By the de Morgan laws, the same is true for $\lor$.
If $(P_n)_{n=1}^\infty$ is a sequence of mutually orthogonal projections of $\cal R_1$, that is, $P_n\leq P_m^\perp$
for all $n\ne m$, then also $E_1(X_n)\leq E_1(X_m)^\perp=E_1(X_m')$. Therefore,
$$
V^*(\bigvee P_n)V=V^*(\sum P_n)V=\sum V^*P_nV=\sum E_1(X_n)=E_1(\bigcup X_n)
$$
(where the series converge weakly)
which shows the $\sigma$-property of $\cal R_1$.
\end{proof}

\begin{corollary}
Let $\Omega_1$  and     $\Omega$ be
complete separable metric spaces and let ${\cal B}(\Omega_1)$ and ${\cal B}(\Omega)$
be their respective Borel $\sigma$-algebras.  Assume that $\Omega_1$ and $\Omega$ 
have the  cardinality  of  $\mathbb R$.
Consider the observables  $E_1:{\cal B}(\Omega_1)\to \lh$ and  $E:{\cal B}(\Omega)\to \lh$  
%for which $E_1({\cal B}(\Omega_1))\subset E({\cal B}(\Omega))$.
such that $E_1$ is a coarse graining of $E$.
If $E_1$ is projection valued, then $E_1=E^f$ for some Borel function $f:\Omega\to\Omega_1$.
\end{corollary}

\begin{proof}
Since  $\Omega_1$ and $\Omega$
are  complete separable metric spaces with the   cardinality of $\mathbb R$, 
according to \cite[Remark (ii), p. 451]{Kuratowski},
there are
bijections 
$\alpha:\Omega \to \mathbb R$ and $\beta:\Omega_1\to\mathbb R$
which are  such that $\alpha,\alpha^{-1},\beta$, and $\beta^{-1}$ are Borel measurable.
 Now $E^\alpha$ and $E_1^\beta$ are real observables with the same ranges as $E_1$
and $E$, respectively. By  \cite[Theorem 3.9]{VSV85}  there is a measurable
function $g:\mathbb R\to \mathbb R$ such that 
$E_1^\alpha(X)=E^\alpha( g^{-1}(X))$, 
$X\in {\cal B}(\mathbb R)$. 
Putting $X={\beta}(Z)$, $Z\in {\cal B}(\Omega)$, we obtain
 $E_1(Z)= E_1^\alpha({\beta}(Z))=
E^\alpha(g^{-1}({\beta}(Z)))=E^f(Z)$, 
 where $f={\beta}^{-1}\circ g\circ\alpha:\Omega \to \Omega_1$.
\end{proof}

\noindent
{\bf Acknowledgement.}
The authors are grateful to Dr. J. Hamhalter, Technical University of Prague, for useful discussions on 
von Neumann algebras. The paper has partially be supported by the grant VEGA No 2/3163/23, Slovak Academy of Sciences,
and by the Science Technology Assistance Agency under contract No APVT-51-032002, Bratislava.


\begin{thebibliography}{References}{}



\bibitem{DP94}
A. Dvure\v censkij, S. Pulmannov\' a,
Difference posets, effects, and a quantum measurement,
{\em Int. J. Theor. Phys.} {\bf 33} (1994) 819-850.

\bibitem{Gudder70}
S. Gudder,
Axiomatic quantum mechanics and generalized probability theory,
{\em Probabilistic Methods in Applied Mathematics},
ed. A.T. Bharucha-Reid, pp. 53-129, 1970.

\bibitem{Halmos}
P.R. Halmos, {\em Measure Theory}, Springer-Verlag, Berlin, 1988. Fourth reprinting of the ed. published by
Van Nostrand, 1950.

\bibitem{Kuratowski}
K. Kuratowski, 
{\em Topology}, vol 1, 
Academic Press, New York, London, 1966.
%Polskie Towarzystwo
%Matematyczne, Warszawa 1962, p. 358.



\bibitem{PLMM92}
P. Lahti, M. M\c aczynski,
Orthomodularity and quadratic transformations in probabilistic theories of physics,
{\em J. Math. Phys.} {\bf 33} (1992) 4133-4138.

\bibitem{PLSP}
P. Lahti, S. Pulmannov\' a,
Coexistence vs. functional coexistence of quantum observables,
{\em Rep. Math. Phys.} {\bf 47} (2001) 199-212.


\bibitem{PLMM95}
P. Lahti, M. M\c aczynski,
Partial order of quantum effects,
{\em J. Math. Phys.} {\bf 36} (1995) 1673-1680.


\bibitem{LY}
P. Lahti, K. Ylinen,
Dilations of positive operator measures and bimeasures related to quantum mechanics,
{\em Math. Slovaca} {\bf 54} (2004) 169-189.

\bibitem{Ozawa}
M. Ozawa,
Phase operator problem and macroscopic extension of quantum mechanics,
{\em Ann. Physics} {\bf 257} (1997) 65-83.

\bibitem{Paulsen}
V.I. Paulsen, {\em Completely bounded maps and dilations}, Longman, Essex, 1986.

\bibitem{Nagy}
F. Riesz, B. Sz.-Nagy, {\em Functional Analysis}, Dover edition 1990, the appendix
originally  published by Ungar Publ. Co. 1960.

\bibitem{Takesaki}
M. Takesaki, {\em Theory of Operator Algebras} I, Springer-Verlag, Berlin, 1979.

\bibitem{VSV85} V.S. Varadarajan,
{\it Geometry of Quantum Theory},
Springer-Verlag, Berlin, 1985.
First edition (in two volumes) by van Nostrand, Princeton, 1968, 1970.


\end{thebibliography}
\end{document}